\documentstyle[epsfig]{mn}

\begin{document}

\title[Triggered Star Formation]{Environmental Dependencies for Star
Formation Triggered by Expanding Shell Collapse}

\author[B. G. Elmegreen, J. Palou\v s and S. Ehlerov\'a]
{B. G. Elmegreen\thanks{E-mail: bge@watson.ibm.com},
J. Palou\v s\thanks{E-mail: palous@ig.cas.cz} and
S. Ehlerov\'a\thanks{E-mail: sona@ig.cas.cz}\\
IBM Research Division, T.J.
Watson Research Center, P.O. Box 218, Yorktown Heights, NY 10598, USA, and\\
Astronomical Institute, Academy of Sciences of the Czech
Republic, Bo\v cn\' \i \ II 1401, 140 31 4, Prague, Czech Republic}

\date{Received 29 October 2001 / Accepted 5 April 2002 }

\maketitle

\markboth{Elmegreen, Palou\v s \& Ehlerov\' a: Triggered Star Formation}{}

\begin{abstract}
Criteria for gravitational collapse of expanding shells in
rotating, shearing galaxy disks were determined using
three-dimensional numerical simulations in the thin shell
approximation. The simulations were run over a grid of 7
independent variables, and the resultant probabilities for
triggering and unstable masses were determined as functions of 8
dimensionless parameters.   When the ratio of the midplane gas
density to the midplane total density is small, an expanding shell
reaches the disk scale height and vents to the halo before it
collapses. When the Toomre instability parameter $Q$, or a similar
shear parameter, $Q_A$, are large, Coriolis forces and shear stall
or reverse the collapse before the shell accumulates enough mass
to be unstable. With large values of $c_{sh}^5/\left(GL\right)$,
for rms velocity dispersion $c_{sh}$ in the swept-up matter and
shell-driving luminosity $L$, the pressure in the accumulated gas is too
large to allow collapse during the expansion time.  Considering
$\sim5000$ models covering a wide range of parameter space, the
common properties of shell collapse as a mechanism for triggered
star formation are: (1) the time scale is $\sim 4\left(c_{sh}/2\pi
G\rho \left[ GL\right] ^{0.2}\right) ^{0.5}$ for ambient midplane
density $\rho $, (2) the total fragment mass is $\sim2 \times 10^7$
M$_\odot$, of which only a small fraction is likely to be
molecular, (3) the triggering radius is $\sim2$ times the scale height,
and the triggering probability is $\sim0.5$ for large OB
associations. Star formation triggered by shell collapse should
be most common in gas-rich galaxies, such as young galaxies or
those with late Hubble types.
\end{abstract}


\section{Introduction}

Giant shells in nearby galaxies often contain young clusters on their
periphery, suggesting that gravitational instabilities in swept-up gas
lead to co-moving, self-gravitating clouds that collapse into stars
(McCray \& Kafatos 1987; Tenorio-Tagle \& Bodenheimer 1988), or
that pre-existing clouds are compressed to an unstable state
as the shell passes, producing the same result (Woodward 1976; Klein et
al. 1985; Dopita, Mathewson \& Ford 1985; Boss 1995; Foster \& Boss
1996; Vanhala \& Cameron 1998; Abrah\'am, Bal\'azs, \& Kun 2000;
Yamaguchi 2001a). 
Sometimes old clusters are found inside
the shells (e.g., Patel et al. 1998; Wilcots \& Miller 1998; Steward
et al. 2000; Stewart \& Walter 2000; Yamaguchi et al. 2001b), in which
case the morphology suggests a sequence of primary and secondary star
formation (see review in Elmegreen 1998).

Star formation that is sequentially triggered like this should be
able to continue for an extended time as one generation leads to
another in any remaining gas. Whether it continues indefinitely
and fills a whole galaxy depends on the triggering efficiency
(Mueller \& Arnett 1976; Gerola \& Seiden 1978). If the mass of
the second generation is larger than the mass of the first, then
the activity should grow until it dominates all star formation.
This may be the case in the solar neighborhood, where most star
formation occurs in dense clusters (Lada, Strom, \& Myers 1993;
Carpenter 2000) that are located at the edges of high pressure
regions, indicative of triggering (e.g., the Trifid Nebula:
Lefloch \& Cernicharo 2000; Rosette: Phelps \& Lada 1997; Orion:
Reipurth, Rodriguez \& Chini 1999; Ophiuchus: de Geus 1992; Sco
Cen: Preibisch \& Zinnecker 1999; Cepheus OB3 and Perseus OB2: 
Sargent 1979; W3/4/5:  Carpenter, Heyer,\&
Snell 2000; see review in Elmegreen et al. 2000). Infrared sources
around the supernova remnant G349.7+0.2 also look triggered
(Reynoso \& Magnum 2000), as do hot cores in the galactic center
cloud Sgr B2 (Martin-Pintado et al. 1999).

Here we investigate the conditions for shell triggering. Numerical
models of expanding, non-magnetic, self-gravitating shells are made,
following the method of Ehlerov\'a \& Palou\v s (1996).  Other models
are in Ehlerov\'a et al. (1997), Efremov et al. (1999) and Ehlerov\' a \&
Palou\v s (2002).  The models use the thin shell approximation proposed
by Sedov (1959), in which a thin, 3-dimensional shell surrounding a hot
medium is divided into a number of elements and a system of equations for
motion, mass and energy is solved.  The approximation is appropriate for
a blastwave propagating in the interstellar medium because the radius
of the shell is much larger than its thickness during the majority of
the evolution.  The main advantage of this method over solutions of the
complete hydrodynamical and Poisson equations in 3 dimensions is a much
lower demand of CPU time.  This gives us the possibility to explore an
extended grid of independent parameters.  However, the adoption of this
method restricts the spatial resolution, which is limited by the finite
thickness of the compressed layer.

\begin{figure}
 \epsfig{figure=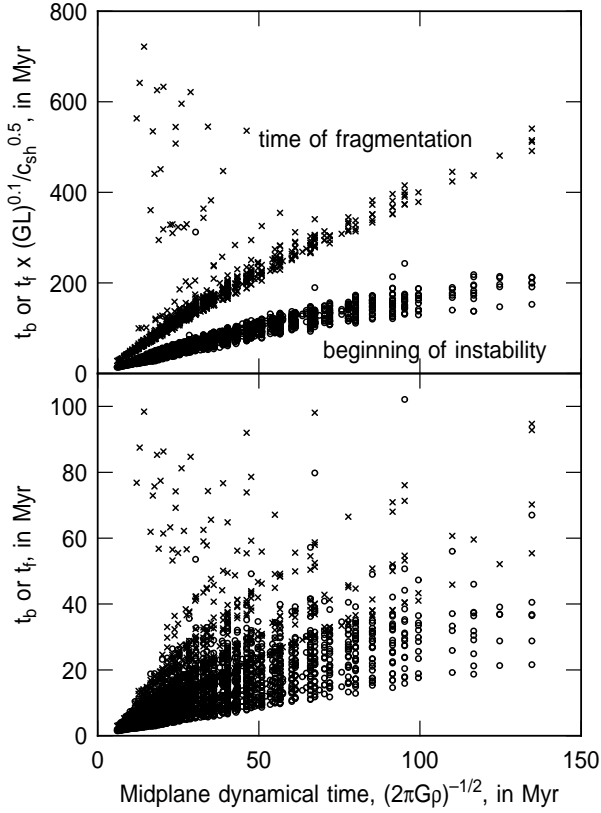,angle=0,width=\hsize}
 \caption{(bottom) The time at the beginning of the
instability, $t_b$ (open circles), and the time of fragmentation, 
$t_f$ (crosses), are
shown versus the dynamical time in the midplane for a variety of
cases with different parameters. 
(top) The time is multiplied by the
dimensionless parameter
$\left(GL\right)^{0.1}/c_{sh}^{0.5}$ for bubble luminosity $L$ and
rms velocity dispersion in the shell, $c_{sh}$. Most collapse
times follow the regular relations given by equations 6 and 7.
Those with very high normalized $t_f$ had high external velocity
dispersions and barely collapsed before they stalled.}
\label{fig1}
\end{figure}
\begin{figure}
\epsfig{figure=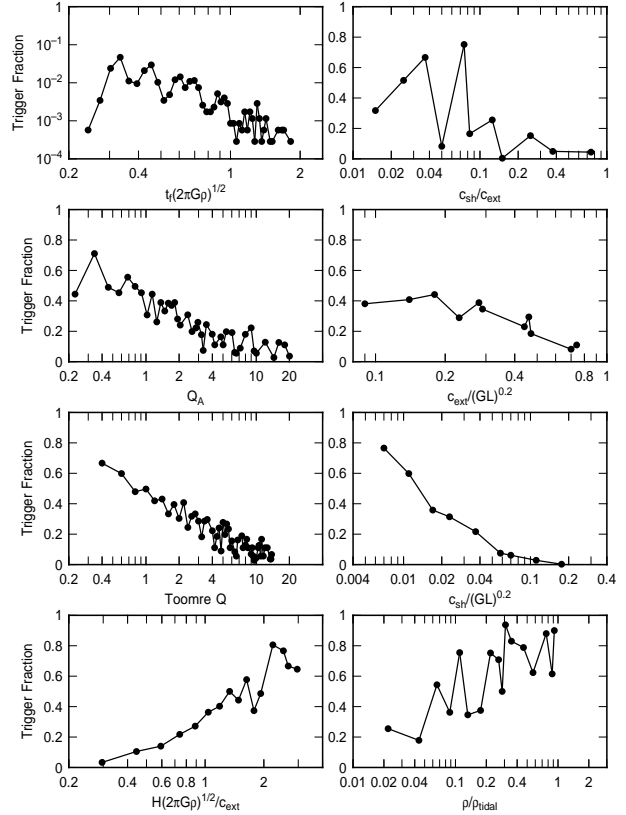,angle=0,width=\hsize}
\caption{The fraction of the models listed in
Table 1 that fragmented before they stalled is plotted versus the
8 dimensionless parameters.  Curves with a strong monotonic
variation indicate that the instability has a sensitive dependence
on that parameter. The most important parameters are the Toomre
parameter, $Q$, and a similar shear parameter, $Q_A$, which
determine the importance of rotation and shear, the density
relative to the critical tidal density, $\rho/\rho_{tidal}$, the
normalized scale height, which is equivalent to the gas mass
fraction in the disk, and the dimensionless rms speed in the
shell, $c_{sh}/\left(GL\right)^{0.2}$.  The figure in the top left
indicates that once the time gets much larger than $\sim0.4$ times
the disk dynamical time, shell collapse becomes unlikely for
typical parameters.}
\end{figure}
\begin{figure}
\epsfig{figure=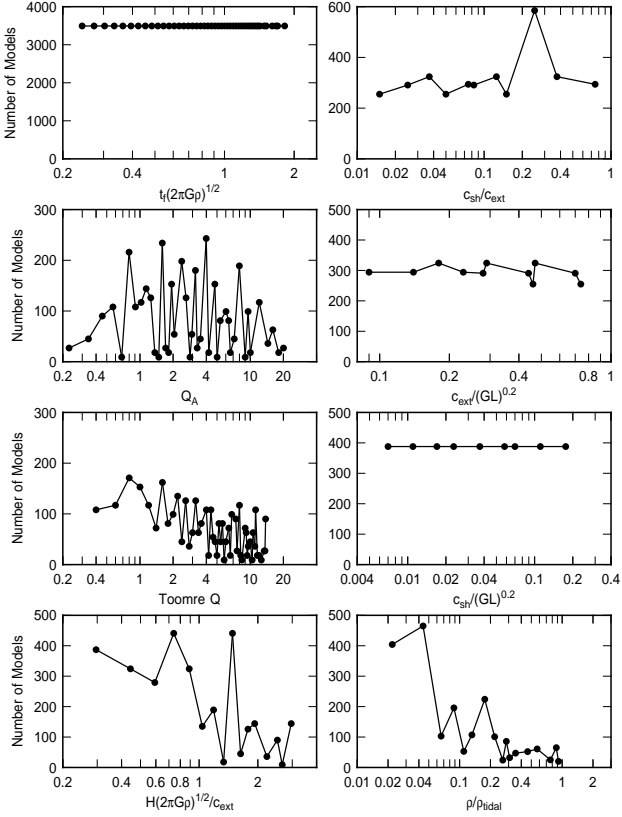,angle=0,width=\hsize}
\caption{The number of models used to make Figure
2 is shown for each dimensionless parameter.}
\end{figure}
\begin{figure}
\epsfig{figure=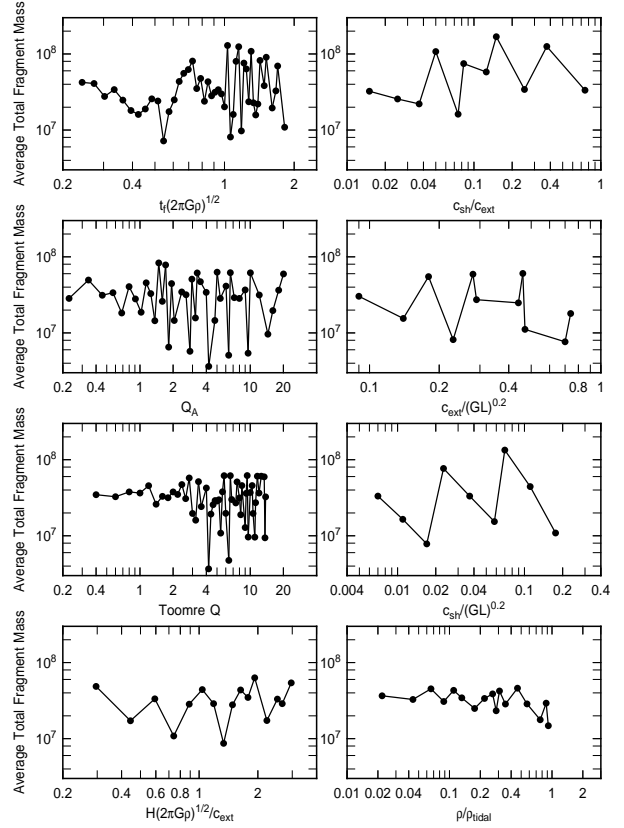,angle=0,width=\hsize}
\caption{The average total fragment mass at the
time of significant collapse, $t_f$,
is shown versus the dimensionless parameters for
each run that was unstable.  The fragment mass does not depend
much on any parameter if the shell goes unstable.}
\end{figure}
\begin{figure}
\epsfig{figure=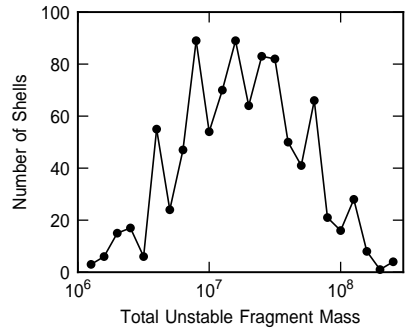,angle=0,width=\hsize}
\caption{A histogram of the total unstable
fragment masses showing the tendency for all unstable shells to
have a fragment mass of a few times $10^7$ M$_\odot$, for a wide range of
ambient densities, scale heights, and other
parameters.}
\end{figure}
\begin{figure}
\epsfig{figure=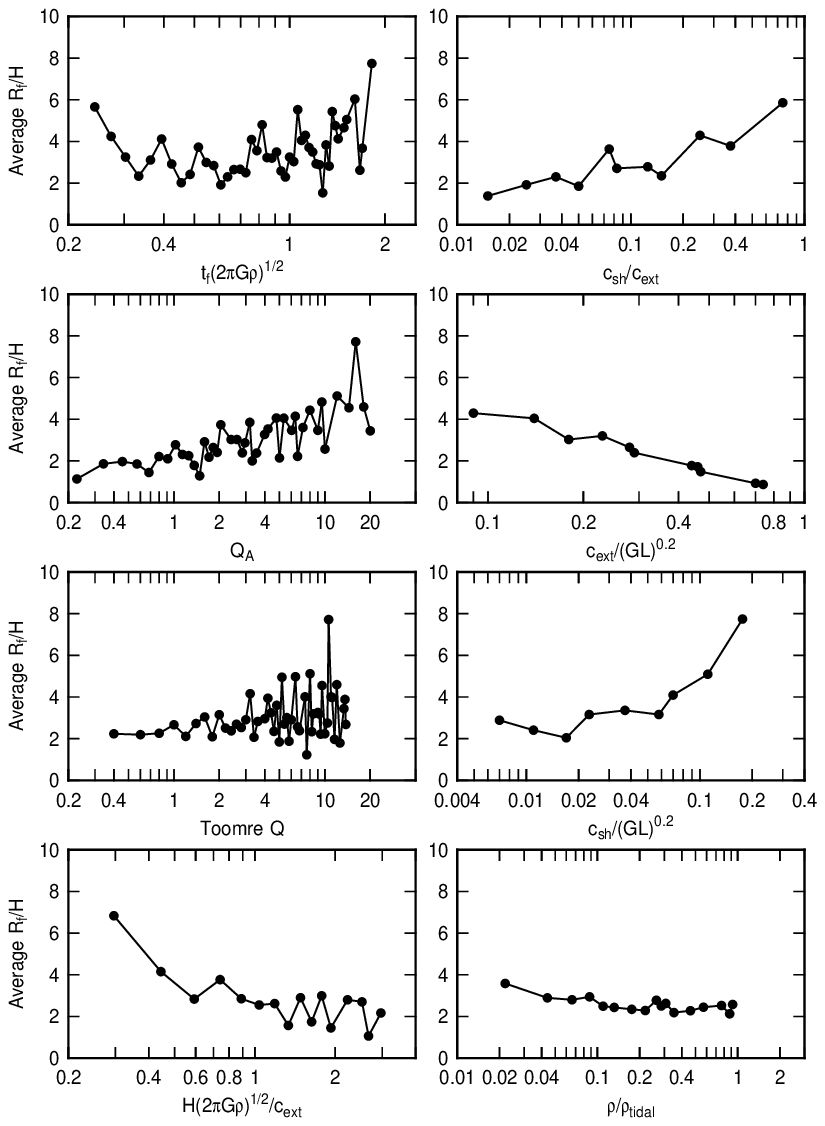,angle=0,width=\hsize}
\caption{The average ratio of the midplane shell
radius to the scale height
at the time of significant fragmentation versus the dimensionless
parameters for each run that
was unstable. This radius is the distance of the most unstable part
of the shell from the expansion center. The shell radius is
always about 2 scale heights when stars begin to form. The dependence on
$c_{ext}/\left(GL\right)^{0.2}$ comes from the theoretical
expression for the expansion of a bubble. The dependence on $Q_A$
comes from the shear that stretches a bubble and increases $R$. }
\end{figure}

\section{Models}

The interstellar medium is modeled by a stratified, non-magnetic
disk with a Gaussian density profile of dispersion $H$, midplane
density $\rho $, velocity dispersion $c_{ext}$, local rotation
curve $V(r)\propto R^\alpha $, and local angular rate, $\Omega $.
Three dimensional expanding shells were driven into this medium by
an energy injection rate $L=E/T$ for energy $E$ and time $T$,
after which the shells coasted on residual internal pressure.
Values of $E=10^{54}$ ergs and $T=1.5 \times 10^7$ yr correspond to
a medium-size OB association.

The expected resistance to expansion by external magnetic
compressive forces can be modeled approximately by increasing
$c_{ext}$ to a value larger than the turbulent speed, perhaps by a
factor of 2. The force from magnetic tension cannot be modeled
this way; it tends to elongate a bubble or contain it within the
disk (Tomisaka 1990, 1998; Ferrier et al. 1991). The implications
of these and other magnetic forces on bubble instability are not
known and will require more elaborate modeling.

The effects of density and velocity irregularities in the swept-up gas
and surrounding medium are not considered (see Silich et al. 1996).
Gravitational instabilities develop faster in the denser parts of this
medium, so the time scale for instabilities and triggering in a clumpy
gas can be shorter than the time scale given here for a uniform gas.
We expect that the clump triggering time can still be estimated from
our results using the derived density dependence for shell collapse at
the higher density of the clump.  The thin shell approximation breaks
down for small dense clumps, however, because they erode by surface
instabilities and get compressed in 3-dimensions when the blastwave
passes (MacLow et al. 1994; Boss 1995; Xu \& Stone 1995; Vanhala, \&
Cameron 1998; Fukuda \& Hanawa 2000).  Our model is not meant to address
this situation.  In a real shell, there may be localized triggering
inside small clouds that are engulfed by the expansion (e.g.,
Yamaguchi et al. 2001a), in addition to large-scale triggering along
the shell periphery where the swept-up gas collapses into new clouds.
Here we consider the latter process.

An expanding shell is approximated by an infinitesimally thin surface
with a local expansion speed $v$, radius $R$, surface density
$\Sigma $, and velocity dispersion, $c_{sh}$. The time $t_b$ at
the beginning of the instability is given by
(Elmegreen 1994)

\begin{equation}
\omega (t_b) = -{{3v}\over{R}}+\left({{v^2} \over {R^2}}+
\left[ {{\pi G\Sigma } \over {c_{sh}}} \right]^2 \right)^{1/2} = 0.
\label{eq:time1}
\end{equation}
For $t > t_b$ and $\omega (t) > 0$,
a fragmentation integral determines the time of significant collapse $t_f$:

\begin{equation}
I_f(t_f)=\int_{t_b}^{t_f}\omega(t^\prime)dt^\prime =1.
\label{eq:time}
\end{equation}
The simulation stops when $v \le c_{ext}$
everywhere in the galaxy symmetry plane. 

The galaxy and shell parameters were varied over the grid of
values in Table 1 to search for trends in the beginning
instability time, the collapse time, and the unstable mass and radius.
The outcome of the expansion depends on more than one
dimensionless variable because there are many competing processes
that can prevent collapse.  Galactic rotation introduces a
Coriolis force that attempts to turn around the expansion of a
bubble and limit its total radius. Galactic shear spreads out the
shell, making an ellipse and concentrating the accumulation at the
tips (Palou\v s, Franco, \& Tenorio-Tagle 1990). The finite scale
height can lead to bubble blowout and depressurization. The
external rms motions lead to erosion when the bubble speed gets
low.  Random sonic and turbulent motions inside the swept-up
matter stabilize it against collapse.

{\tiny
\begin{table*}
\centering
\begin{tabular}{l|l}
\hline
Parameter & Values \\
\hline
Rotation curve slope, $\alpha $ & -0.5, 0, 0.5, 0.8\\
Angular rotation Rate, $\Omega $ & 5, 10, 25, 50, 100 km s$^{-1}$
kpc$^{-1}$\\
Bubble Energy, $E$ & $10^{53}$, $10^{54}$, $10^{55}$ erg\\
Injection Time, $T$ & 15 Myr\\
Disk Scale Height, $H$ & 0.1, 0.2, 0.4, 0.8 kpc\\
Ambient Density, $\rho $ & 0.1, 0.3, 1, 3, 10, 30, 100 H cm$^{-3}$\\
External rms velocity dispersion, $c_{ext}$ & 4, 8, 12, 20 km s$^{-1}$\\
Internal shell velocity dispersion, $c_{sh}$ & 0.3, 1, 3 km s$^{-1}$\\
\hline
\end{tabular}
\caption{Parameter Values}
\label{tab1}
\end{table*}
}

To determine the effects of these processes, we consider
various dimensionless parameters:
\begin{itemize}
\item The Toomre $Q$ parameter for epicyclic
frequency $\kappa=\Omega\left(2\left[1+\alpha\right]\right)^{1/2}$
and disk column density $\Sigma=2\rho H$,
\begin{equation}
Q={{\kappa c_{ext}}\over{\pi G\Sigma}}={{2^{1/2}c_{ext}
\Omega\left(1+\alpha\right)^{1/2}}
\over{2\pi G\rho H}}.
\end{equation}
\item An analogous shear parameter for Oort parameter
$A=-0.5Rd\Omega/dR = 0.5\Omega\left(1-\alpha\right)$,
\begin{equation}
Q_A={{8^{1/2}c_{ext}A}\over{\pi G\Sigma}}={{2^{1/2}c_{ext}
\Omega\left(1-\alpha\right)}\over
{2\pi G\rho H}}.\end{equation}
The constant in this shear-related expression is designed to make $Q_A$
equal to $Q$ for a flat rotation curve ($\alpha=0$).

\item
Another parameter involving gaseous self-gravity is the ratio of
the midplane density to the density at the tidal limit.  The tidal
limiting density for a spherical cluster or cloud that co-rotates
with the galaxy is $\rho_{tidal}=3A\Omega/\left(\pi G\right)$.

\item
The dimensionless scale height for total midplane density $\rho_{total}$,
\begin{equation}
{{H\left(2\pi G\rho\right)^{1/2}}\over{c_{ext}}}
= \left({{\rho}\over{\rho_{total}}}\right)^{1/2}.
\end{equation}
This expression uses $H=c_{ext}/\left(2\pi
G\rho_{total}\right)^{1/2},$ which is valid for an isothermal
self-gravitating layer with $\rho_z(z)=\rho_{total} {\rm
sech}^2\left(z/H\right).$ Our simulations use a Gaussian layer
with $\rho(z)\propto \exp\left(-z^2/H^2\right)$, so the relation
between $\Sigma$, $c$ and $H$ is not exactly the same. The
difference between a Gaussian layer and a ${\rm sech}^2$ layer is
small for $z < 2 H$, so we use the isothermal disk
expressions in this paper.

The dimensionless
scale height is the square root of the ratio of the midplane
gas density to the midplane total density and should be less
than 1.
In the simulations, $H$, $c_{ext},$ and $\rho $ are free parameters and
some combinations give $H\left(2\pi G\rho\right)^{1/2}/c_{ext}$ larger than 1.

\item A dimensionless
parameter that involves the internal rms speed and the shell luminosity is
$c_{sh}/\left(GL\right)^{1/5},$
and the same type of parameter with the external rms speed is
$c_{ext}/\left(GL\right)^{1/5}.$
The first of these luminosity parameters was studied by
Whitworth \& Francis (2002), who found
$c_{sh}/\left(GL\right)^{1/5}<0.13$
as a condition for shell collapse. The second
parameter was studied, in a slightly different form, by
Palou\v s \& Ehlerov\'a (2002), and Ehlerov\'a  \& Palou\v s (2001, 2002).
The ratio of the two rms speeds, $c_{sh}/c_{ext}$, is also considered
here.

\end{itemize}

The total number of parameter combinations from Table 1 is around
20,000, but most of these give unrealistic values for the
dimensionless parameters. We therefore limit the input parameter
values to those which lead to $0.3<Q<15$, $H\left(2\pi
G\rho\right)^{1/2} /c_{ext}<3$, and $\rho/\rho_{tidal}<1$.

\section{Results}

The times for the beginning of the instability, given by equation
\ref{eq:time1}, and the beginning of significant fragmentation,
given by equation \ref{eq:time}, are plotted in Figure 1 versus
the midplane dynamical time, $\left(2\pi G\rho\right)^{-1/2}$. The
range of parameters used for this plot differs from the range in
Table 1 because here we include many different densities. Thus we
consider $Q=0.5,$ 1, 2, and 4 instead of the density explicitly,
and we take $c_{ext}=4$ to 22 km s$^{-1}$ in steps of 2 km
s$^{-1},$ and $c_{sh}=0.5,$ 1, and 3 km s$^{-1}$.  We also fix
$\Omega=25.9$ km s$^{-1}$ kpc $^{-1}$, as in the Solar
neighborhood.  The parameter choices for $H$, $\alpha$, and $L$
are the same as in the table.

Most of the scatter at the bottom of Figure 1 is from variations in shell
rms speed, $c_{sh}$, and bubble luminosity, $L$, which combine into the dimensionless
parameter $c_{sh}^5/\left(GL\right)$. The scatter is reduced at the top of the figure if we
normalize time to $\left(c_{sh}/\left[GL\right]^{0.2}\right)^{0.5}$.
The resulting normalized correlations are
\begin{equation}
t_b\sim 1.5 \left({{c_{sh}/\left[GL\right]^{0.2}}\over{2\pi G\rho}}\right)^{0.5}
\end{equation}
and
\begin{equation}
t_f\sim 4 \left({{c_{sh}/\left[GL\right]^{0.2}}\over{2\pi G\rho}}\right)^{0.5} .
\label{eq:tf}
\end{equation}

The relative importance of the various competing processes is
illustrated in Figure 2, which plots the fraction of the models
in Table 1
that led to collapse versus each dimensionless parameter. The
numbers of models used to determine these fractions are shown in
Figure 3 to illustrate the breadth of coverage of the parameter
space and the likely uncertainty in the triggering fractions.
The top left panel in Figure 2 has a lower triggering fraction than the
other panels because the normalization for each value of the abscissa in
the top left is the constant total number of runs, which is large. In
the other panels, the normalization is the number of runs in each
bin of the dimensionless parameter.  We have to use the total in the
top left because if there is no triggering, then $t_f$ and the abscissa
are not defined.

Figure 2 indicates that gravitational collapse in shearing,
expanding shells is enhanced for large values of $H\left(2\pi
G\rho\right)^{1/2}/c_{ext}$, small values of $Q$ and $Q_A$, large
values of $\rho/\rho_{tidal}$, and small values of
$c_{sh}/\left(GL\right)^{0.2}$. In the first case, the instability
is unlikely for small relative gas densities because the
shell punches through the thin gas layer before it goes
unstable.  The instability is also unlikely for large $Q$ and
$Q_A$ because the corresponding large Coriolis force and shear
slow the expansion, causing the shell to twist around and limit
its accumulation of mass. Collapse is more likely for large
$\rho/\rho_{tidal}$ because the ambient medium is close to
instability anyway. Collapse is more likely for small
$c_{sh}/\left(GL\right)^{0.2}$ because then the shell is thin and
dense for a given radius, so the pressure in the swept-up gas
cannot easily resist its self-gravity.
With $c_{sh}/\left(GL\right)^{0.2}<0.02$ 
for significant triggering, the fragmentation time becomes
$t_f<0.5\left(2\pi G\rho\right)^{-1/2}$, which is less than
half the dynamical time in the external medium.

The density dependence in $t_f$ from equation (\ref{eq:tf}) helps
to explain the $Q$ dependence in Figure 2. If the ambient density
is converted into a mass column density $\Sigma=2\rho H$ using the
velocity dispersion to get the scale height, $H=c_{ext}\left(2\pi
G\rho\right)^{-1/2}$, then
$t_f\sim4\left(c_{sh}/[GL]^{0.2}\right)^{0.5}c_{ext}/\left(\pi
G\Sigma\right)$.  This unstable time, multiplied by $\kappa$, is
$4Q\left(c_{sh}/[GL]^{0.2}\right)^{0.5}$.  Thus large $Q$ or large
$\left(c_{sh}/[GL]^{0.2}\right)^{0.5}$ correspond to an unstable
time longer than an epicycle time, in which case the shell growth
stops by Coriolis forces before it becomes unstable.   If $c_{sh}$
is small because of shell cooling, then the shell can collapse
even if $Q$ is large enough to make the disk otherwise stable.

The total unstable mass in the shell at the fragmentation time $t_f$ 
varies by only a
small amount, a factor of $\sim10$, for all of the unstable
models.  The average of this total mass scales mostly with the
bubble luminosity, but does not correlate with any of the
dimensionless parameters, as shown in Figure 4. Figure 5 plots a
histogram of the number of shell models in regularly spaced
logarithmic intervals of the total unstable mass.  The shells need
a sufficiently large mass to be unstable for typical expansion
speeds.  The total mass involved is $\sim2\times10^7$ M$_\odot$,
although the molecular part that forms stars will be smaller.

The shell radius at the fragmentation time, $R_f$, is shown relative
to the scale height, $H$, in Figure 6.  This radius is defined to be the
distance from the expansion center, measured along the plane, of the most
unstable part of the shell at the time of fragmentation.  
The ratio $R_f/H$ depends mostly on the internal sound
speed, aside from the usual dependence on $c_{ext}/\left(GL\right)^{0.2}$
that comes into the theoretical expression for the bubble expansion
(Weaver et al. 1977).  There is a slight correlation between $R_f/H$
and $Q_A$, but not with $Q$, because shear stretches a bubble and gives it
a larger average radius at the fragmentation time.  The results suggest
that the shell radius along the plane is $\sim2$ to 3 scale heights
when star formation begins. This distance corresponds to $\sim200$
to 300
pc for typical disks, in agreement with observations. The shell
size should be $\sim2$ times larger in dwarf galaxies because of their
larger scale heights (e.g., Walter \& Brinks 1999).  A shell is larger
than this perpendicular to the plane because of its more rapid expansion
into lower densities.

\section{Conclusions}

Sequential star formation triggered by the collapse of expanding
shells requires essentially three conditions: that the midplane
density of the gas be comparable to the stellar midplane density,
that the Toomre parameter, $Q$, be relatively low, and that the
velocity dispersion in the swept-up gas be relatively low compared
to $\left(GL\right)^{0.2}$. The first of these conditions avoids
blow out above and below the plane in the time it takes the shell
to go unstable. The second avoids shell stalling by Coriolis
forces, and the third makes the shell dense and strongly
self-gravitating (Whitworth \& Francis 2002). 
Other dimensionless parameters have a similar
sensitivity: $Q_A$ and the inverse of $\rho/\rho_{tidal}$ behave
like $Q$.

When and where are the dimensionless conditions for triggering
fulfilled? In early type spiral galaxies where the relative gas
density is low, $Q$, $Q_A$ and $\rho_{tidal}/\rho$ can be large
(Caldwell et al. 1992) and $H\left(2\pi
G\rho\right)^{1/2}/c_{ext}$ can be small. In this case, shell
instabilities require very small values of $c_{sh}/(G L)^{0.2}$;
i.e., either very large shell-driving luminosities or very low rms
dispersions and temperatures in the swept-up gas. This is rather
exceptional, although perhaps the triggering conditions are
fulfilled in the vicinity of OB 78 in M31. In late type spirals
and gas rich dwarf galaxies, $Q$, $Q_A$, and $\rho_{tidal}/\rho$
are small and $H\left(2\pi G\rho\right)^{1/2}/c_{ext}$ is large.
Consequently, triggered star formation in expanding shells should
be more frequent, as in the supergiant shells of the LMC
(Yamaguchi et al. 2001a,b) and IC 2574 (Stewart \& Walter 2000).

The relative gas density is often high in starburst regions too, but
there the absolute gas density is high as well, so the interstellar
dynamical time can be very short. If it is shorter than the lifetime of
an O-type star, which is typical for nuclear starburst regions,
then bubble expansions will not be significantly aided by supernova
explosions before the collapse time. This lowers $L$ for a given OB
association mass.  Starburst regions can also have large $c_{ext}$,
which limits shell growth and makes HII regions ineffective at moving
gas around.  As a result, spontaneous star formation processes, or those
triggered by random supernovae, are more likely.

Perhaps the best place for shell-triggered star formation was in
the early Universe, when most galaxies were gas-rich and the
dynamical time for an average galactocentric radius was much
longer than an O-star lifetime.  Galaxies at high redshift should
contain many gaseous shells and triggering events, causing them to
resemble sheared and massive versions of today's irregular dwarf
galaxies.

Acknowledgements:  Helpful comments by the referee are gratefully
acknowledged. Support for BGE came from NSF grant AST-9870112.  JP and
SE acknowledge financial support by the Grant Agency of the Academy of
Sciences of the Czech Republic under the grants No.  A3003705, B3003106
and K1048102.

\end{document}